\documentclass[onecolumn]{elsart}
\usepackage{natbib}

\usepackage{graphicx}
\usepackage{times}

\begin{document}

\begin{frontmatter}



\title{The influence of forward-scattered light in transmission measurements of (exo)planetary atmospheres}


\author[sron]{R.J.~de~Kok}, and
\author[sron]{D.M.~Stam}

\address[sron]{SRON Netherlands Institute for Space Research, 
               Sorbonnelaan 2, 3584 CA Utrecht, the Netherlands}
			
%% This copyright statement isn't required at any stage by the Icarus
%% Editorial Office or Elsevier.  However, until you sign over the
%% copyright to Elsevier prior to publication (or negotiate with them
%% about copyright), you own the copyright to anything you create.
%% Just to keep things unambiguous, always include a copyright statement
%% or explicitly dedicate your work to the public domain.
\begin{center}
\scriptsize
%Copyright \copyright\ 2006 
\end{center}

%% --- ELSEVIER STUFF ---
%% The commands below up to the \end{frontmatter} are commented out
%% so that we can do some Icarus-required formatting on the second and
%% third pages that is not required later on by Elsevier.  So when
%% your paper gets accepted, and you are ready to start dealing with
%% Elsevier, copy your abstract and keywords up here, uncomment these
%% lines, and comment out the ICARUS STUFF below.
%% 
%% Alternately, you might just want to move these abstract, keyword,
%% and end frontmatter commands down, and comment out the ICARUS STUFF
%% commands.  It doesn't matter.

% \begin{abstract}
% % Text of abstract

% 
% \end{abstract}
% 
% \begin{keyword}
% % keywords here, in the form: keyword \sep keyword
% 
% 
% % PACS codes here, in the form: \PACS code \sep code
% 
% \end{keyword}

%% --- END ELSEVIER STUFF ---

\end{frontmatter}

%% --- ICARUS STUFF ---
%% Some formatting on the first, second, and third pages are required
%% by the Icarus Editorial Office that are not required by Elsevier.
%% This section contains those things.  When you are ready to transition
%% to ``Elsevier'' mode, copy your abstract and keywords up into
%% the ELSEVIER STUFF section, and then you can just delete everything
%% in this section.

%% We need to list the number of manuscript pages, figures, and tables. 
%%
%% Rather than manually count these things out, we'll use a little
%% trick here from Paul.  All you have to do is place three \label{}
%% tags on the last page, the last table, and the last figure, that
%% way these values are automatically updated (as long as you remember
%% to move the lasttable and lastfig labels when you add or remove
%% tables and figures).

\begin{flushleft}
\vspace{1cm}
%Number of pages: 30 \\
%Number of tables: 0 \\
%Number of figures: 6 \\
\end{flushleft}

%% Don't worry about finding the various last* tags and deleting them
%% when you go to ``Elsevier'' mode if you don't want to, they should be
%% silently ignored.

%% Icarus wants keywords separated by a semicolon.
\newcommand{\sep}{; }

%% The second page should indicate a proposed running head of not more 
%% than 55 characters, and the name and address to which editorial 
%% correspondence and proofs should be directed.

\pagebreak

\noindent
\textbf{Proposed Running Head:}\\
  Light scattering in planet transmission measurements
%          1         2         3         4         5
% 1234567890123456789012345678901234567890123456789012345

\vspace{3cm}
\noindent
\textbf{Please send Editorial Correspondence to:} \\
Remco de Kok \\
SRON Netherlands Institute for Space Research \\
Sorbonnelaan 2\\
3584 CA Utrecht \\
The Netherlands \\
\\
Email: R.J.de.Kok@sron.nl\\
Phone: +31 887775725 \\
Fax: +31 887775601

\vfill

\pagebreak

%% We can't really use the elsart ``abstract'' environment because
%% for submission, Icarus wants the Running Head and Editorial
%% Correspondence stuff on the second page.  So we just have to
%% cook up some formatting that sets off the abstract text on the
%% third page.

\noindent
\textbf{ABSTRACT}

The transmission of light through a planetary atmosphere can be studied as 
a function of altitude and wavelength using stellar or solar occultations, 
giving often unique constraints on the atmospheric composition. For exoplanets, 
a transit yields a limb-integrated, wavelength-dependent transmission spectrum 
of an atmosphere. When scattering haze and/or cloud particles are present in 
the planetary atmosphere, the amount of transmitted flux not only depends on 
the total optical thickness of the slant light path that is probed, but also
on the amount of forward-scattering by the scattering particles. Here, we 
present results of calculations with a three-dimensional Monte Carlo code 
that simulates the transmitted flux during occultations or transits. For 
isotropically scattering particles, like gas molecules, the transmitted flux 
appears to be well-described by the total atmospheric optical thickness. 
Strongly forward-scattering particles, however, such as commonly found in 
atmospheres of Solar System planets, can increase the transmitted flux significantly.
For exoplanets, such added flux can decrease the apparent radius of the planet 
by several scale heights, which is comparable to predicted and measured features 
in exoplanet transit spectra. We performed detailed calculations for Titan's 
atmosphere between 2.0~and 2.8~$\mu$m and show that haze and gas abundances 
will be underestimated by about 8\% if forward-scattering is ignored in the
retrievals. 
At shorter wavelengths, errors in the gas and haze abundances and in the 
spectral slope of the haze particles can be several tens of percent, also for 
other Solar System planetary atmospheres. We also find that the contribution
of forward-scattering can be fairly well described by modelling the atmosphere
as a plane-parallel slab. This potentially reduces the need for a full 
three-dimensional Monte Carlo code for calculating transmission spectra 
of atmospheres that contain forward-scattering particles.

% %% Keywords should appear after the abstract. 
\vspace{\fill}

\pagebreak
%% --- END ICARUS STUFF ---

%%%%%%%%%%%%%%%%%%%%%%%%%%%%%%%%%%%%%%%%%%%%%%%%%%%%%%%%%%%%%%%%%%%%%%%%%%%%%%%%%%%%%%%

\section{Introduction}

Planetary atmospheres can be studied remotely using spectroscopy of the light that 
is reflected or the radiation that is emitted by the planet. Another method for
studying planetary atmospheres is to measure the light of a bright source, such 
as a star or the Sun, as it is attenuated by (part of) the atmosphere. For Solar 
System objects, such transmission measurements are performed during stellar or 
solar occultations, when the planetary limb is in between the light source and 
the spacecraft or telescope. One advantage of transmission measurements is that 
the light source is often very bright, enabling high signal-to-noise measurements.
Another advantage is that transmission measurements allow sampling of long path 
lengths through the atmosphere, since the light travels through the curved limb 
of the planet. Long path lengths increase the sensitivity to e.g.~trace gases 
with very small concentrations. Furthermore, altitudes in an atmosphere 
can be probed with a high vertical resolution, as only a very limited altitude range is probed by a single observation compared to e.g.~on-disc observations, and many altitudes can be sampled due to the high signal-to-noise. Also, occultation measurements rely on relative changes in the measured signal, and not on absolute signal levels, and are hence `self-calibrating' and less dependent on instrument drifts. As a result of these advantages, transmission 
measurements are very suitable for retrieving vertical profiles of trace gases. 
Disadvantages of solar and/or stellar occultations are that they require very
special observing geometries, making it not trivial to target specific places on the planet. Furthermore, only a single location on the planet can be studied 
simultaneously, and the long path lengths prevent probing the lower, thicker 
layers of a planetary atmosphere. For solar occultations, the atmosphere can naturally only be studied at local twilight conditions.
In recent years there have been many occultation measurements of Solar System 
planets. Besides the Earth, occultation transmission measurements have been 
performed of Venus \citep[e.g.][]{van08,fed08,bel12}, Mars 
\citep[e.g.][]{bla89,kra89,for09}, Jupiter 
\citep{for03}, Saturn \citep{kim12}, Titan \citep{bel09,kim11}, and 
Pluto \citep{hub88,ell03}, for which more are planned \citep{ste08}. 
Besides measuring the atmospheric transmission, occultations can also 
be used to measure the atmospheric refraction, which we will not consider
here.

%%%
Transmission measurements are one of the most valuable probes of the 
atmospheres of extrasolar planets (exoplanets). The transmission of an 
exoplanetary atmosphere can be measured when the orbit of the exoplanet with
respect to the observer is such that the planet crosses the disc of its
host star. 
By accurately measuring the wavelength-dependence of the decrease of 
star light during such a transit, a transmission spectrum of the planetary
atmosphere can be derived \citep{sea00,bro01,hub01}, although only the day-night boundary can be probed. These transit measurements 
have revolutionised our knowledge of exoplanet atmospheres and are now widely 
used to derive atmospheric properties of a number of transiting exoplanets. There are now several planets for which molecular absorption features have been identified through measuring their transits \citep[e.g.][]{cha02,tin07,sne10,sea10}, as well as molecular and haze Rayleigh scattering \citep[e.g.][]{lec08,pon08,sin11}. Also, there is an indication of high wind speeds on the planet HD 209458b \citep{sne10}. The smallest planet for which there have been significant transit measurements is the `super-Earth' GJ 1214b \citep{cha09}. Transit measurements allow the determination of the bulk composition of this planet \citep{mil10}, but at present it is not clear whether it is has a water vapour atmosphere, or a hydrogen atmosphere with clouds \citep{bea10,bea11,cro11,dem12}. Unfortunately, atmospheres with high molecular weight, like Earth, are hard to characterise in this way, as the spectral features are much reduced compared to hydrogen atmospheres. This is because the scale height of the atmosphere scales with the molecular weight.

The transmission of a planetary atmosphere is mainly determined by the 
extinction or total optical thickness of gases and particles, like haze and 
cloud particles, measured along the line of sight.
The total optical thickness $\tau_{\rm t}$ can be split into an absorbing part,
$\tau_{\rm a}$, and a scattering part, $\tau_{\rm s}$, using the 
single-scattering albedo $\tilde{\omega}$ of the mixture of gases
and particles in the layer:
\begin{equation}
   \tau_{\rm t}= \tilde{\omega}\tau_{\rm t} + (1-\tilde{\omega}) \tau_{\rm t}=
   \tau_{\rm s} + \tau_{\rm a}.
\end{equation}
The optical thicknesses $\tau_{\rm t}$, $\tau_{\rm a}$, and $\tau_{\rm s}$
and the single scattering albedo $\tilde{\omega}$
are usually wavelength dependent.

If the transmission $T$, i.e. the ratio of transmitted to incident flux, is purely 
determined by the extinction along the line of sight, it is simply given by the Beer-Lambert law:
\begin{equation}
   T= e^{-\tau_{\rm t}}
\label{eq.transm}
\end{equation}
If the scattering optical thickness $\tau_{\rm s}$ along the line of sight is 
larger than zero, however, a fraction of the light that has been scattered 
out of the incident beam of light, will be scattered in exactly the forward 
direction after a single scattering or after multiple scattering events. 
This means that more light reaches the observer than expected from the
extinction optical thickness of the atmosphere alone. In fact, large particles 
scatter light predominantly in the forward direction because of diffraction \citep[see e.g.][]{han74}, 
possibly making the forward-scattered light contribute significantly to the total 
transmitted flux. Indeed, non-absorbing, perfectly forward-scattering particles 
would appear completely transparent in transmission, despite a non-zero optical 
thickness. 

\begin{figure}[htp]
\centering
\includegraphics[width=0.9\textwidth]{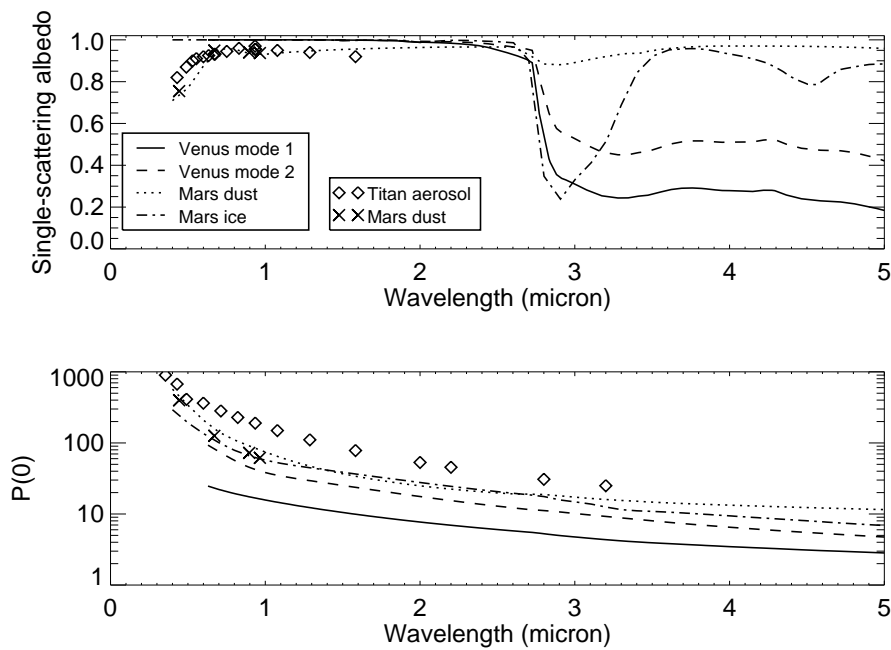}

\caption{Single-scattering albedos $\tilde{\omega}$ and phase functions in 
the forward direction, $P(0)$, of particles present in the atmospheres
of Venus, Mars and Titan (see the text for details).}
\label{fig.ssa+p0}
\end{figure}

How a particle scatters the incident flux as a function of the scattering angle 
$\Theta$ is described by the phase function $P(\Theta)$. 
Figure~\ref{fig.ssa+p0} shows the single-scattering albedo $\tilde{\omega}$ 
and the phase function in the forward direction, $P(0)$, of some particles
found in atmospheres of Solar System planets.
The phase functions have been normalised such that their integral over 4$\pi$ 
steradians is unity. We show measured as well as calculated values of $P(0)$.
The properties of Titan aerosol and martian dust particles are derived from 
measurements by \citet{tom08} and \citet{tom99}, respectively. The lines 
in Fig.~\ref{fig.ssa+p0} are computed using 
Mie-theory \citep[][]{der84}. 
Note that the strength of the forward-scattering part of the phase function 
is due to diffraction and as such depends mostly on the size of the particles, 
and less on their shape \citep[e.g.][]{mis96}. Hence, Mie scattering should be accurate for the purpose of estimating the forward-scattering peak.
For the Venus cloud particles, we used the optical constants of \citet{pal75} 
and the size distributions for the two most dominant size modes from \citet{gri93}. 
For the calculations for martian dust and ice cloud particles we used the optical 
parameters and size distributions as described by \citet{kle09}. It can be seen 
that especially at short wavelengths $P(0)$, and hence the forward-scattering 
contribution to the transmission, can potentially be large, even though all these 
different particles are relatively small (with effective radii smaller than 2 $\mu$m).
The forward-scattering part of the phase function of larger particles, such as
liquid water cloud particles found on Earth, can be much larger
\citep[see e.g. Fig.~11.13.6 in][]{mis06}. 

Analysing transmission measurements using Eq.~\ref{eq.transm} and thus ignoring the 
(potential) contribution of forward-scattered flux to the transmission, which is how these
measurements are usually analysed, will result in an error in the retrieved optical 
thickness of the planetary atmosphere. 
In this paper, we explore the contribution of the forward-scattered flux to the 
transmission signal using a three-dimensional Monte Carlo model, which simulates 
scattered light in a spherical, stratified atmosphere. Besides providing general results, 
we will also present more detailed results for Titan's atmosphere, to serve as a 
practical example. Finally, we draw conclusions regarding possible errors in retrievals 
from transmission measurements when scattering particles are present.

%%%%%%%%%%%%%%%%%%%%%%%%%%%%%%%%%%%%%%%%%%%%%%%%%%%%%%%%%%%%%%%%%%%%%%%%%%%%%%%%%%%%%%%

\section{Numerical model}

Our model is a Monte Carlo model, whose core is based on subroutines of the \emph{mc-unik} code by Andreas Macke \citep[pers.~comm.;][]{mac99,cah05}.
In a Monte Carlo model, the light that is incident on the planetary atmosphere 
is described by a large number of separate photons, the paths of which are followed
through the atmosphere. In our model, photons are fired through the atmosphere 
from one side and in one direction, and their locations and directions when they 
leave the atmosphere (provided they have not been absorbed) are stored.
We ignore refraction and polarisation. Refraction will generally increase the path length through the atmosphere, but its effect at low pressures has been found to be small by \citet{hub01} and \citet{bel09}.
A model atmosphere is assumed to be spherically symmetric and is described by an 
arbitrary number of shells. In our calculations, the thickness of each atmospheric shell is roughly a sixth of the atmospheric scale height. We did not try to find the optimum for this thickness, but halving the number of layers did not significantly change our scattering results. Also considering our simulations with a single layer, which show that the spherical geometry is not important to first order for the scattering contribution (see Sections 3 and 4), we do not expect large errors in the scattering contribution due to our finite number of layers (of the order of 50). When analysing real measurements, more layers might be needed, depening on the required model accuracy. Across each atmospheric shell, the
temperature, pressure, gas and particle concentrations and their optical 
properties are assumed to be constant. In the results presented below, we always have only one type of scatterer present in the atmosphere. The atmosphere is bounded below
by a black surface at a pressure of 5 bar. Generally, the slant optical thickness of a light path crossing the 5 bar level is of the order of 100 or more and hence the light emanating from this level has a negligible contribution to the transmitted signal.

When a photon enters an atmospheric shell (either the outer or an inner shell,
and either from the outside or the inside), the free path length $l^*$ of 
the photon in the shell is calculated using:
\begin{equation}
   l^* = \tau_{\rm t}^* \frac{dl}{d\tau_{\rm t}},
\end{equation}
where $\tau_{\rm t}^*$ follows from
\begin{equation}
   e^{-\tau^*_{\rm t}} = \xi,
\end{equation}
with $\xi$ a random number between 0 and 1. 
Within each shell in our atmosphere, $d\tau_{\rm t}/dl$ is constant. 

If the free path length and the direction of the photon are such that 
the photon passes through the shell without being intercepted by a gas
molecule or a particle, we stop the photon where it would leave the 
inner or outer boundary of the atmospheric shell. 
If the photon is leaving the outer atmospheric shell towards space, 
its location and direction of propagation are stored and a new photon 
is fired. 
If the photon is about to enter another atmospheric shell, we calculate
the photon's free path length in the new shell and let it continue
its path in the new shell.

If the free path length and the direction of a photon in an atmospheric
shell are such that the photon is intercepted by a gas molecule or a 
particle in the same shell, we calculate whether the photon is absorbed
or scattered, based on the single-scattering albedo $\tilde{\omega}$
of the mixture of gas molecules and particles in the shell. 
If the photon is absorbed, it is lost and a new photon is fired.
If the photon is scattered, its direction changes according to the phase
function $P$ of the mixture of gas molecules and particles in the shell.
The photon's scattering angle $\Theta^*$, and hence its new direction, 
is determined by the cumulative distribution
of the phase function as follows:
\begin{equation}
   \int_0^{\Theta^*} P(\Theta^*) d\Theta^* = 
   \xi \int_{0}^{\pi} P(\Theta) d\Theta,
\end{equation}
with $\xi$ a random number between 0 and 1.
For this new direction, a new free path length $l^*$ is calculated
and the photon is sent on its way.

\begin{figure}[htp]
\centering
\includegraphics[width=0.9\textwidth]{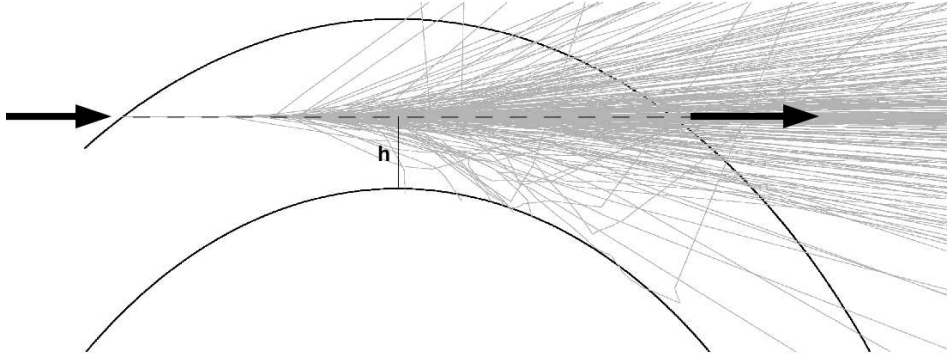}

\caption{Illustration of the geometries. The illumination source is on the left
         side of the planet, and the observer on the right side. 
         Photons are fired in a single direction on the left, indicated by the 
         arrow on the left. The observer measures the photons that leave
         the atmosphere in the direction of the arrow on the right.
         The solid lines indicate the top and bottom of the atmosphere in the plane of the emission and observer and the tangent altitude of the observation is denoted by $h$. Apparent penetration of the rays at the bottom of the atmosphere is caused by the rays moving in the dimension perpendicular to the figure.}
 \label{fig.mc}
\end{figure}

A photon reaching the surface below the atmosphere is assumed to be absorbed. 
So, all photons are either absorbed or leave the top of the atmosphere. 
Figure~\ref{fig.mc} shows a sketch of the geometries involved for photons fired at a tangent altitude $h$. The figure 
includes sample trajectories of photons calculated using our Monte Carlo model.
For exoplanets, illumination from all values of $h$, reaching from the bottom to the top of the atmosphere, has to be taken into account. In our model, we find that the contribution of photons exiting at a different tangent altitude from where the photons are inserted, in exactly the right direction, is negligible.

We have tested our Monte Carlo code using an atmosphere consisting of a plane-parallel
slab, for which several accurate codes for calculating scattered fluxes exist. 
We compared our calculations of the total transmitted flux and the flux
transmitted in a specific direction for isotropic scattering with \citet{shet70} 
and \citet{deh87} (ignoring polarisation), and for non-isotropic scattering 
with \citet{deh87} and \citet{hov85}. In all cases, the results agree to within a 
percent. We have also tested our code assuming a spherical shell model atmosphere 
containing only purely absorbing particles. For such an atmosphere, the number of 
photons transmitted through the atmosphere, divided by the total number of fired 
photons, should simply equal the transmission $T$ as described by Eq.~\ref{eq.transm}, 
which indeed it does to well within a percent. Hence, our sampling of the mean free 
paths is also accurate in the three-dimensional model.

For every tangent altitude $h$, we fire a number of photons that is equal to $10^5$ 
divided by the transmission derived from the total optical thickness along the 
line-of-sight (Eq.~\ref{eq.transm}), to ensure that enough photons survive their 
trip through the atmosphere to have relatively accurate results. Afterwards, we
normalize the number of surviving photons to the number of incident photons
for each tangent altitude $h$.
We impose a maximum number of photons of $10^8$ for large optical thicknesses, 
to keep the computation time manageable. This means that, for large optical depths, less than 10$^5$ photons are transmitted and the accuracy of the result decreases.

%%%%%%%%%%%%%%%%%%%%%%%%%%%%%%%%%%%%%%%%%%%%%%%%%%%%%%%%%%%%%%%%%%%%%%%%%%%%%%%%%%%%%%%

\section{Scattering results and application to exoplanets}

We have performed calculations of the flux that is transmitted through a planetary 
atmosphere and that is detected at the other side, with the transmitted photons 
travelling in the same direction as the incident ones. In this section, we 
consider three model atmospheres that differ in the type of scattering particles: 
Rayleigh scattering particles, and two types of 
forward-scattering particles, with different forward-scattering strengths. As briefly discussed in the introduction, one would qualitatively expect more photons to reach the observer if scattering is included, with more forward-scattering particles giving a larger added signal. 
In each model atmosphere, the particles are uniformly mixed throughout
the atmosphere in hydrostatic equilibrium and they are the only opacity source.
The three model atmospheres have the same extinction optical thickness,
and are based on the `hot Jupiter' HD 189733b, 
for which Rayleigh-scattering particles have been identified in the atmosphere 
\citep{pon08,sin11}. Although we present here only the results with parameters from one specific atmosphere, it will be clear later that these results are also valid for other planets. A uniform temperature of 1000~K and a mean molecular weight 
of 2.2 gram/mol is assumed, with a gravitational acceleration of 18 m/s$^2$. 

The forward-scattering particles have Henyey-Greenstein phase functions
\citep[][]{hen41},
given by
\begin{equation}
\label{eq.henyey}
     P(\Theta) = \frac{1}{4\pi} \frac{1 - g^2}{(1 + g^2 - 2g \cos \Theta)^{3/2}},
\end{equation}
with $g$ the asymmetry parameter. For one type of particles, we chose $g=0.9$,
which represents a case of moderate forward scattering, with $P(0)=15.1$ 
(cf.~Fig.~\ref{fig.ssa+p0}). The other type of particles is more 
forward-scattering, with $g=0.98$, corresponding to $P(0)=393.8$.
All particles are assumed to be non-absorbing, with single-scattering 
albedos equal to unity.

\begin{figure}[htp]
\centering
\includegraphics[width=0.9\textwidth]{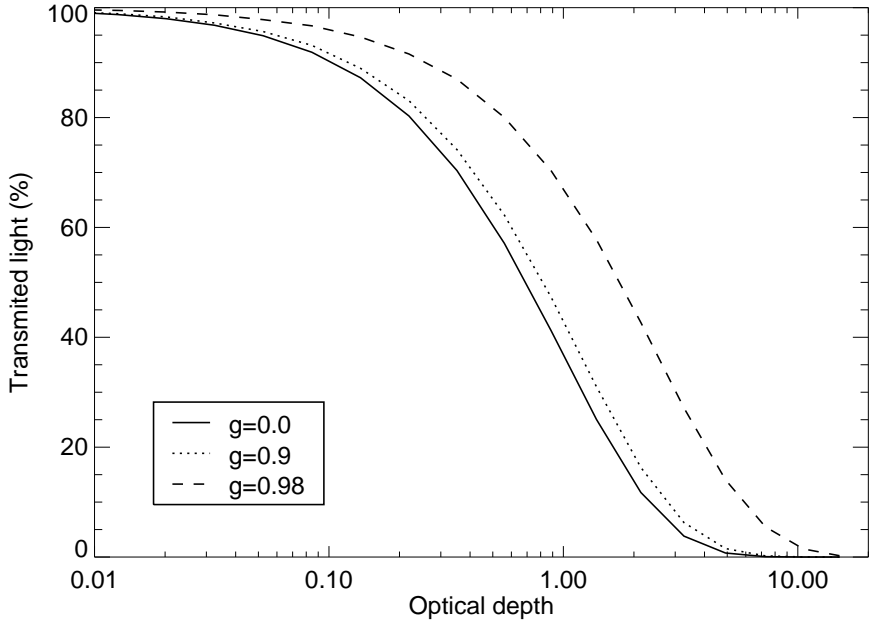}

\caption{The transmitted flux as a function of the slant optical thickness 
         through the atmosphere for isotropically scattering particles ($g=0.0$) 
         and two types of Henyey-Greenstein
         scattering particles ($g=0.9$ and $g=0.98$). The line for isotropically 
         scattering particles is practically indistinguishable from a line (not shown) 
         representing transmission without added forward-scattering scattering 
         (Eq.~\ref{eq.transm}).}
	 \label{fig.transm}
\end{figure}

Figure~\ref{fig.transm} shows the transmitted flux for different tangent heights 
$h$ (see Fig.~\ref{fig.mc}), plotted as a function of the total optical thickness 
along the line of sight for the three model atmospheres.
In the case of Rayleigh scattering particles, only very little extra 
flux is transmitted as a result of the forward-scattering and the total 
transmitted flux can be well described by Eq.~\ref{eq.transm}. Indeed, adding
the forward-scattered flux to the transmission changes the transmitted flux
by less than a percent (relatively speaking, so not in percentage points; i.e.~ a 1\% change of a transmission of 20\% gives 19.8\% here, not 19\%) for 
optical thicknesses less than unity, and by a few percent for optical 
thicknesses around~10. 
In the cases with the forward-scattering particles, however, the 
forward-scattering adds significantly to the transmitted flux as expected
based on extinction alone (see Eq.~\ref{eq.transm} and the line for $g=0.0$ in 
Fig.~\ref{fig.transm}).
Especially at large optical thicknesses, these scattering particles can yield 
a transmitted flux that is several times the flux that would be transmitted 
if the particles were purely absorbing. For $g=0.98$ this is most apparent, but also for $g=0.9$ this is the case for very low transmission values, which are not well visible in the figure. These results confirm our qualitative expectations, as noted in the introduction, but now the magnitude of the effect is made apparent.
In terms of absolute number of photons, optical thicknesses around unity give the 
largest differences between forward-scattering and purely absorbing particles. 
From our code, we also know the number of times a photon is scattered before it 
exits the atmosphere at the top or the bottom. As expected, for small optical
thicknesses, the additional scattering contribution comes mostly from photons 
that are scattered only once, whereas for large optical thicknesses multiple 
scattering is more important.

To simulate the transmission for transiting exoplanets, the transmitted flux
must be integrated along the planetary limb that is in front of the star.
By measuring the transmitted flux at a given wavelength, one can determine 
the apparent radius of the planet at that wavelength \citep[e.g.][]{bro01}.
Our calculations show that the apparent radius of exoplanet HD 189733b with an
atmosphere filled with forward-scattering particles, decreases
0.7\% and 3.8\% for $g=0.9$ and $g=0.98$, respectively. This corresponds to 
roughly 1 and 4 times the atmospheric scale height and a decrease in transit depth 
of 0.03 and 0.15 percentage points (from an initial 2\% transit depth). 
Such differences are easily comparable to the depth of spectral features in 
the transit spectrum, and to differences in measured transit depth 
\citep[e.g.][]{sin11}. For other exoplanets, the number in terms of scale height will be similar, as we found by changing gravity.

We also compared our Monte Carlo results with calculations of light passing 
through a plane-parallel slab \citep[][]{deh87} with the same scattering 
particles and the same total optical thickness as the light path in the 
three-dimensional, spherical atmosphere. 
For all tangent heights and corresponding optical thicknesses, the relative difference between the flux transmitted through the
three-dimensional atmosphere and that transmitted through the plane-parallel 
slab calculations is less than 10\%, meaning that plane-parallel slab 
calculations \citep[e.g.][]{hub01} can describe the effects of forward-scattering
particles fairly well. This is perhaps not surprising, since by far the largest
optical thickness along the line of sight is located at a relatively small region 
in the atmosphere, around the tangent point, for each (almost) parallel beam that is passing through the atmosphere, from the star to the observer.

The outcome of this last test suggests that the exact geometry is not important to first order, which means that Fig.~3 is valid for all different kinds of atmospheres, if there is no absorption. Different atmospheres will then only differ in the amount of slant optical thickness as a function of height and wavelength. The case when there is also absorption by e.g.~gases present is illustrated in the next section.

%%%%%%%%%%%%%%%%%%%%%%%%%%%%%%%%%%%%%%%%%%%%%%%%%%%%%%%%%%%%%%%%%%%%%%%%%%%%%%%%%%%%%%%

\section{Application to Titan}

To further explore the effect that scattering haze and/or cloud particles have on 
transmitted fluxes and the atmospheric information derived from transmission measurements,
we take Titan's atmosphere as example, for which a sequence of near-infrared solar occultation 
measurements by Cassini VIMS were analysed by \citet{bel09}. The presence of gas absorption bands makes this case more complex. Again, we expect scattering from the haze to increase the amount of observed photons. In gas absorption bands this effect is expected to be less important, but it is not directly apparent what the magnitude of the effect is and what effect the scattering will have on retrieved gas abundances. 
\citet{bel09} derive the optical thickness and spectral slope of Titan's haze
and gas abundances, all as functions of altitude, from these observations at a single location.
Here, we do not attempt to fit the VIMS data, but instead simulate transmission 
spectra using our Monte Carlo code and fully including forward-scattering particles. 
From these simulated spectra, we then derive gas and haze properties ignoring 
scattering, like \citet{bel09} do. We then compare the retrieved atmospheric 
parameters with the `real' parameters that we put into our code to quantify 
potential errors in retrievals when forward-scattering is ignored.

For Titan's atmosphere we calculated transmission spectra between 2.0 and 2.8~$\mu$m 
with a spectral resolution of 16.6~nm \citep{bro04} using the correlated-$k$ 
method \citep{lac91} and CH$_4$ line parameters from HITRAN 2008 \citep{rot09}. Opacity distribution tables are calculated for the ranges of temperatures and pressures relevant for Titan's atmosphere.
Since we are investigating the effects of ignoring forward-scattering and because
we are not fitting real measurements, we don't have to be concerned about potential 
errors in the gas absorption coefficients due to the low temperatures in Titan's
atmosphere; we use the same gas absorption properties in the calculations that 
include forward-scattering as in the calculations that ignore this scattering. Similarly, we are not concerned about small errors expected from the correlated-$k$ approximation, especially since by far the most opacity along the light path crossing the atmosphere is located in a single atmospheric layer, whose altitude is the tangent height. Potentially uncorrelated $k$-distributions at different altitudes will therefore have little effect.
We assumed a pressure-temperature profile from \citet{vin06} and a uniform 
CH$_4$ Volume Mixing Ratio (VMR) of 1.4\% \citep{nie05}. The haze optical thickness
as a function of altitude at 2 $\mu$m was taken from \citet{bel09} and the spectral 
slope of the haze is taken to be that of the upper haze of \citet{tom08}. The 
single-scattering albedo of the haze particles as a function of wavelength was 
linearly extrapolated from \citet{tom08} and the phase functions were linearly 
interpolated from the phase functions tabulated by \citet{tom08} (see 
Fig.~\ref{fig.ssa+p0}). For our Monte Carlo calculations we took into account 
the distance of the spacecraft to Titan \citep{bel09} as well as the VIMS 
telescope dimensions \citep{bro04} to collect all of the `observed' photons. How the photons are imaged after hitting the primary mirror is not taken into account, but this is expected to have very little influence, because by far the largest contribution of the signal comes from the direction of the Sun.

\begin{figure}[htp]
\centering
\includegraphics[width=0.9\textwidth]{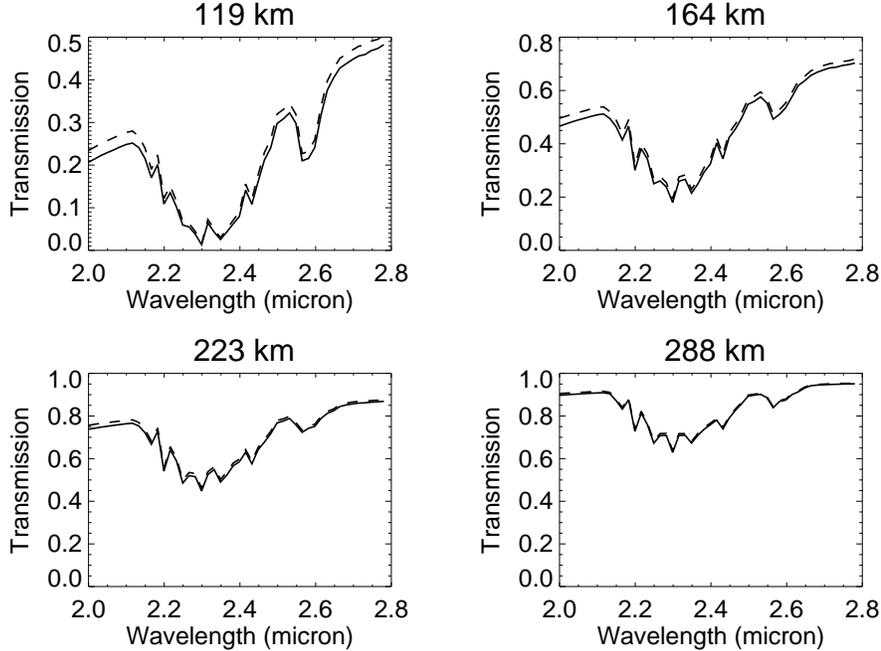}

\caption{Calculated transmission spectra with (dashed lines) and without 
         (solid lines) forward-scattering for Titan's atmosphere, for four
         different observed tangent altitudes (see Fig.~\ref{fig.mc}.)}
	 \label{fig.titanspec}
\end{figure}

Figure~\ref{fig.titanspec} shows calculated transmission spectra with 
and without including the forward-scattering contribution for different
observer's tangent altitudes. This part of the 
spectra shows a CH$_4$ absorption feature and \citet{bel09} derive the 
CH$_4$ VMR from this part of the spectrum (cf.~their Fig.~7). From our figure,
it is clear that the contribution of forward-scattering by the haze particles
is largest in the continuum, where there is little gas absorption, as expected. 
The added contribution of forward-scattering to the continuum flux makes the 
absorption features deeper with respect to the continuum. It also slightly flattens the slope of the continuum 
flux, because the haze particles are more forward-scattering at smaller wavelengths
than at larger wavelengths.

We derived the atmospheric parameters by fitting the transmission spectra 
that we calculated using the Monte Carlo code, which includes scattering. These spectra (dashed lines in Fig.~4) are thus thought of as the `real' measurements that need fitting. For the fitting, we ignored 
the scattering, and used just Eq.~\ref{eq.transm}, similar to what \citet{bel09} do. The retrieved atmospheric parameters can then be compared to the `real' state of the atmosphere, as input in our scattering model.
We first scaled the amount of haze particles and the spectral slope 
(the Angstr\"om exponent) of the haze to exactly match the beginning (2~$\mu$m) 
and end (2.8~$\mu$m) of the transmission spectra. 
Compared to the 'real' values that we used in the calculation of the
transmission spectra, ignoring the forward-scattering in the fit scales 
down the density of haze particles by typically 8\% and the Angstr\"om exponent
by 3\%. Then, we used a least-squares fit to determine the CH$_4$ VMR. 
Figure~6 shows the retrieved CH$_4$ VMR as a function of altitude
together with the 'real' VMR.
The noise on the retrieved values is the result of the
limited number of photons used in the Monte Carlo calculations.
The decrease of the error in the retrieved VMR with altitude above about 200~km,
corresponds with the decrease of the difference between the transmission spectra
in Fig.~\ref{fig.titanspec}. Figure~6 shows that in this case, 
ignoring forward-scattering when retrieving the gas abundance can lead to 
an underestimation of the gas VMR of typically 8\%. 

The reason for the underestimation is that ignoring the added continuum flux 
that is due to forward-scattering leads to a lower retrieved haze particle 
density (since more flux is transmitted in the continuum). It is not straightforward to argue what effect this will have on the depth of the absorption band compared to the `real' spectrum, since the scattering contribution to the transmitted signal is not related to the optical depth in a simple way as the non-scattering signal (Eq.~2) is. The outcome of lowering the haze opacity in the non-scattering case is illustrated in Fig.~5. The dotted line shows the fit when the haze is scaled down, but the methane abundance is kept identical. The apparent increase in depth of the absorption band between the solid (no forward scattering) and dashed (with forward scattering) spectra corresponds simply to an almost uniform lowering of the optical depth with wavelength, giving rise to a non-uniform increase in transmission according to Eq.~2. However, whether this increase of band depth is more or less than the increase of band depth obtained from the scattering contribution in the dashed spectrum is not immediately clear. It can be seen from Fig.~5 that in our calculations, the absorption band is too deep for the non-scattering case (dotted line) and the CH$_4$ abundance should be decreased to fit the dashed line. 
Surprisingly perhaps, ignoring forward-scattering thus leads to an \emph{under}estimation of both 
the gas and the haze abundances if measurements are fitted. 
It is therefore all the more puzzling that \citet{bel09} find a CH$_4$ abundance that is significantly larger than that measured by the Huygens probe \citep{nie05} 
below 200 km. \citet{bel09} suggest various possible explanations for this effect, 
such as extra absorption within the haze itself.

\begin{figure}[htp]
\centering
\includegraphics[width=0.9\textwidth]{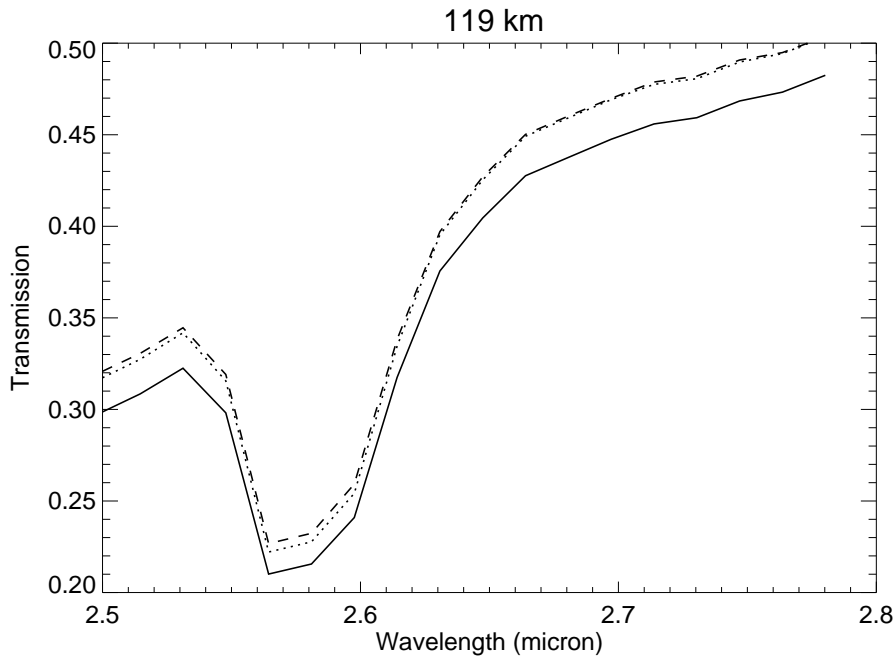}
\caption{A zoom of a spectrum with dashed and solid lines as in Fig.~4. The dotted line now shows the absorption-only fit to the spectrum when the haze opacity and its spectral slope is decreased, but the methane abundance is kept identical.}
\end{figure}

\begin{figure}[htp]
\centering
\includegraphics[width=0.9\textwidth]{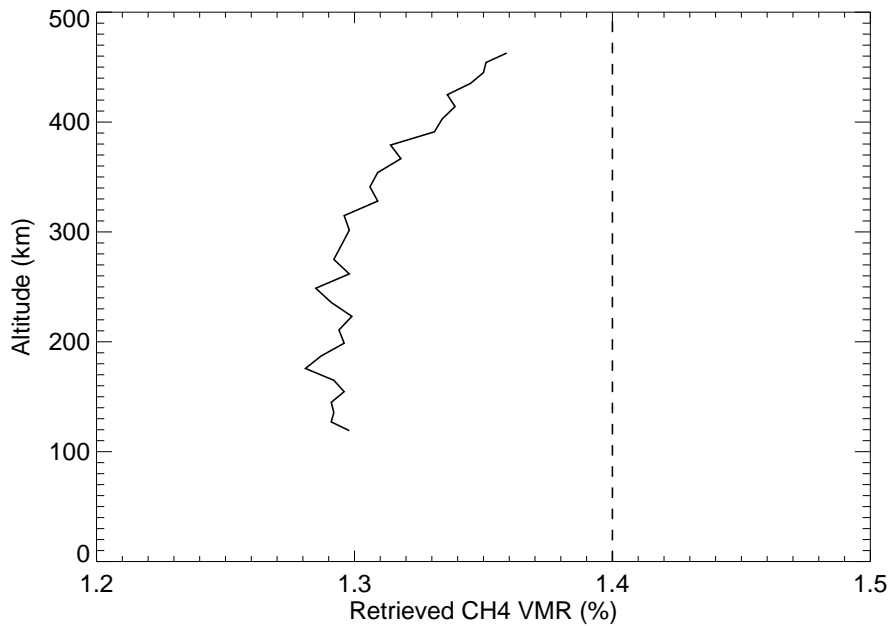}

\caption{The CH$_4$ VMR as retrieved while ignoring forward-scattering by the haze
         particles (solid line). The vertical dashed line indicates the 'real' VMR 
         value that we used as input for our spectral calculations, which included 
         forward-scattering.}
	 \label{fig.vmrfit}
\end{figure}

As we did for the exoplanet calculation (see Section 3), we compared the transmission spectra as calculated with our Monte Carlo code
and assuming a spherical model atmosphere with spectra calculated assuming a 
plane-parallel slab model atmosphere with the same particle scattering properties, weighted by the slant 
optical thickness of each atmospheric layer, 
and the same optical thickness (as measured along a slant light path for a 
given tangent altitude). The differences between the spherical atmosphere 
spectra and the plane-parallel slab spectra are small: of the order of a 
few percent in the transmitted flux.

%%%%%%%%%%%%%%%%%%%%%%%%%%%%%%%%%%%%%%%%%%%%%%%%%%%%%%%%%%%%%%%%%%%%%%%%%%%%%%%%%%%%%%%

\section{Conclusions and discussion}

We have developed a three-dimensional, spherical Monte Carlo code to simulate 
the added contribution of scattered light to the flux that is transmitted through
a planetary atmosphere during stellar or solar occultations and exoplanet transits. 
Our calculations with parameters of the exoplanet HD 189733b show that Rayleigh scattering does not add 
significantly to the transmitted flux that one would expect when the transmission 
depended solely on the extinction (absorption plus scattering) optical thickness 
of the atmosphere. However, forward-scattering particles can contribute 
significantly to the transmitted flux.
For (slant) optical thicknesses larger than unity, the added contribution of 
forward-scattered flux can be as large or larger than the flux expected from 
the extinction alone. For exoplanet transits, this means that the magnitude of 
a transit of a planet with forward-scattering particles in its atmosphere
can be several percent smaller than predicted when these particles are assumed to be fully absorbing. Such a difference in transit depth is 
easily comparable to the magnitude of measured and predicted spectral features 
in a transit spectrum.

With our Monte Carlo code, we performed detailed calculations of occultation 
spectra of Titan's atmosphere between 2.0 and 2.8 $\mu$m and show that the 
haze abundance and trace gas mixing ratios are underestimated by $\sim 8$\% 
if the forward-scattering contribution to the transmitted flux is ignored
in the retrieval algorithm. Also in such a case, the derived spectral slope 
of the haze particles is underestimated by a few percent. Figure~\ref{fig.ssa+p0} 
shows that Titan's haze particles are fairly forward-scattering between 2~and
3$\mu$m, but that cloud and haze particles can be at least an order of magnitude 
more forward-scattering at shorter wavelengths. This means that at shorter
wavelengths, cloud, haze or gas abundances could be underestimated by several 
tens of percent if forward-scattering is ignored. Figure~\ref{fig.ssa+p0} 
also shows that such strong forward-scattering is not uncommon in other 
atmospheres in the Solar System. Also the change of forward-scattering with 
wavelength can be much stronger at shorter wavelength, resulting in much 
larger errors in the derived spectral slope of forward-scattering particles for absorption-only models. Similarly, fast changes of the single-scattering albedo with wavelength can also cause significant errors in the derived spectral slope 
of cloud and haze particles.
 
We also compared our Monte Carlo calculations for three-dimensional, spherical
planetary atmospheres to calculations for atmospheres composed of plane-parallel 
slabs, with the same optical thickness as the slant light paths through the 
spherical atmosphere, and with scattering properties weighted by the slant 
optical thickness of each atmospheric layer (see e.g.~\citet{hub01,kim11} 
for examples).  For the cases we considered, we found these two calculations to match 
within 10\% for optical thicknesses less than 10. This has the implication that 
the contribution of forward-scattered light to transmitted fluxes can be fairly 
well modelled using a plane-parallel assumption, for which radiative transfer
routines are available that are much faster than a three-dimensional Monte Carlo 
code.

%%%%%%%%%%%%%%%%%%%%%%%%%%%%%%%%%%%%%%%%%%%%%%%%%%%%%%%%%%%%%%%%%%%%%%%%%%%%%%%%%%%%%%%

\textbf{Acknowledgements}

We acknowledge financial support by the Netherlands Organisation for Scientific 
Research (NWO). We thank Andreas Macke for use of part of his code. We thank the two anonymous reviewers for their careful reading and useful suggestions.

%%%%%%%%%%%%%%%%%%%%%%%%%%%%%%%%%%%%%%%%%%%%%%%%%%%%%%%%%%%%%%%%%%%%%%%%%%%%%%%%%%%%%%%
\clearpage

%\bibliographystyle{elsart-harv}
%\bibliography{/home/remcodk/texfiles/references3}

\begin{thebibliography}{47}
\expandafter\ifx\csname natexlab\endcsname\relax\def\natexlab#1{#1}\fi
\expandafter\ifx\csname url\endcsname\relax
  \def\url#1{\texttt{#1}}\fi
\expandafter\ifx\csname urlprefix\endcsname\relax\def\urlprefix{URL }\fi

\bibitem[{{Bean} et~al.(2011){Bean}, {D{\'e}sert}, {Kabath}, {Stalder},
  {Seager}, {Miller-Ricci Kempton}, {Berta}, {Homeier}, {Walsh}, and
  {Seifahrt}}]{bea11}
{Bean}, J.~L., {D{\'e}sert}, J.-M., {Kabath}, P., {Stalder}, B., {Seager}, S.,
  {Miller-Ricci Kempton}, E., {Berta}, Z.~K., {Homeier}, D., {Walsh}, S.,
  {Seifahrt}, A., 2011. {The Optical and Near-infrared Transmission Spectrum of
  the Super-Earth GJ 1214b: Further Evidence for a Metal-rich Atmosphere}.
  Astrophys. J. 743, 92.

\bibitem[{{Bean} et~al.(2010){Bean}, {Miller-Ricci Kempton}, and
  {Homeier}}]{bea10}
{Bean}, J.~L., {Miller-Ricci Kempton}, E., {Homeier}, D., 2010. {A ground-based
  transmission spectrum of the super-Earth exoplanet GJ 1214b} 468, 669--672.

\bibitem[{{Bellucci} et~al.(2009){Bellucci}, {Sicardy}, {Drossart}, {Rannou},
  {Nicholson}, {Hedman}, {Baines}, and {Burrati}}]{bel09}
{Bellucci}, A., {Sicardy}, B., {Drossart}, P., {Rannou}, P., {Nicholson},
  P.~D., {Hedman}, M., {Baines}, K.~H., {Burrati}, B., 2009. {Titan solar
  occultation observed by Cassini/VIMS: Gas absorption and constraints on
  aerosol composition}. Icarus 201, 198--216.
  
\bibitem[{{Belyaev} et~al.(2012){Belyaev}, {Montmessin}, {Bertaux}, {Mahieux},
  {Fedorova}, {Korablev}, {Marcq}, {Yung}, and {Zhang}}]{bel12}
{Belyaev}, D.~A., {Montmessin}, F., {Bertaux}, J.-L., {Mahieux}, A.,
  {Fedorova}, A.~A., {Korablev}, O.~I., {Marcq}, E., {Yung}, Y.~L., {Zhang},
  X., 2012. {Vertical profiling of SO$_{2}$ and SO above Venus' clouds by
  SPICAV/SOIR solar occultations}. Icarus 217, 740--751.
  
\bibitem[{{Blamont} et~al.(1989){Blamont}, {Chassefiere}, {Goutail}, {Mege},
  and {Nunes-Pinharanda}}]{bla89}
{Blamont}, J.~E., {Chassefiere}, E., {Goutail}, J.~P., {Mege}, B.,
  {Nunes-Pinharanda}, M., 1989. {Vertical profiles of dust and ozone in the
  Martian atmosphere deduced from solar occultation measurements}. Nature 341,
  600--603.

\bibitem[{{Brown} et~al.(2004){Brown}, {Baines}, {Bellucci}, {Bibring},
  {Buratti}, {Capaccioni}, {Cerroni}, {Clark}, {Coradini}, {Cruikshank},
  {Drossart}, {Formisano}, {Jaumann}, {Langevin}, {Matson}, {McCord},
  {Mennella}, {Miller}, {Nelson}, {Nicholson}, {Sicardy}, and {Sotin}}]{bro04}
{Brown}, R.~H., {Baines}, K.~H., {Bellucci}, G., {Bibring}, J.-P., {Buratti},
  B.~J., {Capaccioni}, F., {Cerroni}, P., {Clark}, R.~N., {Coradini}, A.,
  {Cruikshank}, D.~P., {Drossart}, P., {Formisano}, V., {Jaumann}, R.,
  {Langevin}, Y., {Matson}, D.~L., {McCord}, T.~B., {Mennella}, V., {Miller},
  E., {Nelson}, R.~M., {Nicholson}, P.~D., {Sicardy}, B., {Sotin}, C., 2004.
  {The Cassini Visual And Infrared Mapping Spectrometer (Vims) Investigation}.
  Space Science Reviews 115, 111--168.

\bibitem[{{Brown}(2001)}]{bro01}
{Brown}, T.~M., 2001. {Transmission Spectra as Diagnostics of Extrasolar Giant
  Planet Atmospheres}. Astrophys. J. 553, 1006--1026.

\bibitem[{{Cahalan} et~al.(2005){Cahalan}, {Oreopoulos}, {Marshak}, {Evans},
  {Davis}, {Pincus}, {Yetzer}, {Mayer}, {Davies}, {Ackerman}, {Barker},
  {Clothiaux}, {Ellingson}, {Garay}, {Kassianov}, {Kinne}, {Macke}, {O'Hirok},
  {Partain}, {Prigarin}, {Rublev}, {Stephens}, {Szczap}, {Takara},
  {V{\'a}rnai}, {Wen}, and {Zhuravleva}}]{cah05}
{Cahalan}, R.~F., {Oreopoulos}, L., {Marshak}, A., {Evans}, K.~F., {Davis},
  A.~B., {Pincus}, R., {Yetzer}, K.~H., {Mayer}, B., {Davies}, R., {Ackerman},
  T.~P., {Barker}, H.~W., {Clothiaux}, E.~E., {Ellingson}, R.~G., {Garay},
  M.~J., {Kassianov}, E., {Kinne}, S., {Macke}, A., {O'Hirok}, W., {Partain},
  P.~T., {Prigarin}, S.~M., {Rublev}, A.~N., {Stephens}, G.~L., {Szczap}, F.,
  {Takara}, E.~E., {V{\'a}rnai}, T., {Wen}, G., {Zhuravleva}, T.~B., 2005. {THE
  I3RC: Bringing Together the Most Advanced Radiative Transfer Tools for Cloudy
  Atmospheres.} Bulletin of the American Meteorological Society 86, 1275--1293.

\bibitem[{{Charbonneau} et~al.(2009){Charbonneau}, {Berta}, {Irwin}, {Burke},
  {Nutzman}, {Buchhave}, {Lovis}, {Bonfils}, {Latham}, {Udry}, {Murray-Clay},
  {Holman}, {Falco}, {Winn}, {Queloz}, {Pepe}, {Mayor}, {Delfosse}, and
  {Forveille}}]{cha09}
{Charbonneau}, D., {Berta}, Z.~K., {Irwin}, J., {Burke}, C.~J., {Nutzman}, P.,
  {Buchhave}, L.~A., {Lovis}, C., {Bonfils}, X., {Latham}, D.~W., {Udry}, S.,
  {Murray-Clay}, R.~A., {Holman}, M.~J., {Falco}, E.~E., {Winn}, J.~N.,
  {Queloz}, D., {Pepe}, F., {Mayor}, M., {Delfosse}, X., {Forveille}, T., 2009.
  {A super-Earth transiting a nearby low-mass star} 462, 891--894.

\bibitem[{{Charbonneau} et~al.(2002){Charbonneau}, {Brown}, {Noyes}, and
  {Gilliland}}]{cha02}
{Charbonneau}, D., {Brown}, T.~M., {Noyes}, R.~W., {Gilliland}, R.~L., 2002.
  {Detection of an Extrasolar Planet Atmosphere}. Astrophys. J. 568, 377--384.

\bibitem[{{Croll} et~al.(2011){Croll}, {Albert}, {Jayawardhana}, {Miller-Ricci
  Kempton}, {Fortney}, {Murray}, and {Neilson}}]{cro11}
{Croll}, B., {Albert}, L., {Jayawardhana}, R., {Miller-Ricci Kempton}, E.,
  {Fortney}, J.~J., {Murray}, N., {Neilson}, H., 2011. {Broadband Transmission
  Spectroscopy of the Super-Earth GJ 1214b Suggests a Low Mean Molecular Weight
  Atmosphere}. Astrophys. J. 736, 78.

\bibitem[{{de Haan} et~al.(1987){de Haan}, {Bosma}, and {Hovenier}}]{deh87}
{de Haan}, J.~F., {Bosma}, P.~B., {Hovenier}, J.~W., 1987. {The adding method
  for multiple scattering calculations of polarized light}. Astron. Astrophys.
  183, 371--391.

\bibitem[{{de Mooij} et~al.(2012){de Mooij}, {Brogi}, {de Kok}, {Koppenhoefer},
  {Nefs}, {Snellen}, {Greiner}, {Hanse}, {Heinsbroek}, {Lee}, and {van der
  Werf}}]{dem12}
{de Mooij}, E.~J.~W., {Brogi}, M., {de Kok}, R.~J., {Koppenhoefer}, J., {Nefs},
  S.~V., {Snellen}, I.~A.~G., {Greiner}, J., {Hanse}, J., {Heinsbroek}, R.~C.,
  {Lee}, C.~H., {van der Werf}, P.~P., 2012. {Optical to near-infrared transit
  observations of super-Earth GJ 1214b: water-world or mini-Neptune?} 538, A46.

\bibitem[{{de Rooij} and {van der Stap}(1984)}]{der84}
{de Rooij}, W.~A., {van der Stap}, C.~C.~A.~H., 1984. {Expansion of Mie
  scattering matrices in generalized spherical functions}. Astron. Astrophys.
  131, 237--248.

\bibitem[{{Elliot} et~al.(2003){Elliot}, {Ates}, {Babcock}, {Bosh}, {Buie},
  {Clancy}, {Dunham}, {Eikenberry}, {Hall}, {Kern}, {Leggett}, {Levine},
  {Moon}, {Olkin}, {Osip}, {Pasachoff}, {Penprase}, {Person}, {Qu}, {Rayner},
  {Roberts}, {Salyk}, {Souza}, {Stone}, {Taylor}, {Tholen}, {Thomas-Osip},
  {Ticehurst}, and {Wasserman}}]{ell03}
{Elliot}, J.~L., {Ates}, A., {Babcock}, B.~A., {Bosh}, A.~S., {Buie}, M.~W.,
  {Clancy}, K.~B., {Dunham}, E.~W., {Eikenberry}, S.~S., {Hall}, D.~T., {Kern},
  S.~D., {Leggett}, S.~K., {Levine}, S.~E., {Moon}, D.-S., {Olkin}, C.~B.,
  {Osip}, D.~J., {Pasachoff}, J.~M., {Penprase}, B.~E., {Person}, M.~J., {Qu},
  S., {Rayner}, J.~T., {Roberts}, L.~C., {Salyk}, C.~V., {Souza}, S.~P.,
  {Stone}, R.~C., {Taylor}, B.~W., {Tholen}, D.~J., {Thomas-Osip}, J.~E.,
  {Ticehurst}, D.~R., {Wasserman}, L.~H., 2003. {The recent expansion of
  Pluto's atmosphere} 424, 165--168.

\bibitem[{{Fedorova} et~al.(2008){Fedorova}, {Korablev}, {Vandaele}, {Bertaux},
  {Belyaev}, {Mahieux}, {Neefs}, {Wilquet}, {Drummond}, {Montmessin}, and
  {Villard}}]{fed08}
{Fedorova}, A., {Korablev}, O., {Vandaele}, A., {Bertaux}, J., {Belyaev}, D.,
  {Mahieux}, A., {Neefs}, E., {Wilquet}, W.~V., {Drummond}, R., {Montmessin},
  F., {Villard}, E., 2008. {HDO and H$_{2}$O vertical distributions and
  isotopic ratio in the Venus mesosphere by Solar Occultation at Infrared
  spectrometer on board Venus Express}. Journal of Geophysical Research
  (Planets) 113, doi:10.1029/2008JE003146.

\bibitem[{{Forget} et~al.(2009){Forget}, {Montmessin}, {Bertaux},
  {Gonz{\'a}lez-Galindo}, {Lebonnois}, {Qu{\'e}merais}, {Reberac},
  {Dimarellis}, and {L{\'o}pez-Valverde}}]{for09}
{Forget}, F., {Montmessin}, F., {Bertaux}, J.-L., {Gonz{\'a}lez-Galindo}, F.,
  {Lebonnois}, S., {Qu{\'e}merais}, E., {Reberac}, A., {Dimarellis}, E.,
  {L{\'o}pez-Valverde}, M.~A., 2009. {Density and temperatures of the upper
  Martian atmosphere measured by stellar occultations with Mars Express
  SPICAM}. Journal of Geophysical Research (Planets) 114, 1004.

\bibitem[{{Formisano} et~al.(2003){Formisano}, {D'Aversa}, {Bellucci},
  {Baines}, {Bibring}, {Brown}, {Buratti}, {Capaccioni}, {Cerroni}, {Clark},
  {Coradini}, {Cruikshank}, {Drossart}, {Jaumann}, {Langevin}, {Matson},
  {McCord}, {Mennella}, {Nelson}, {Nicholson}, {Sicardy}, {Sotin},
  {Chamberlain}, {Hansen}, {Hibbitts}, {Showalter}, and {Filacchione}}]{for03}
{Formisano}, V., {D'Aversa}, E., {Bellucci}, G., {Baines}, K.~H., {Bibring},
  J.~P., {Brown}, R.~H., {Buratti}, B.~J., {Capaccioni}, F., {Cerroni}, P.,
  {Clark}, R.~N., {Coradini}, A., {Cruikshank}, D.~P., {Drossart}, P.,
  {Jaumann}, R., {Langevin}, Y., {Matson}, D.~L., {McCord}, T.~B., {Mennella},
  V., {Nelson}, R.~M., {Nicholson}, P.~D., {Sicardy}, B., {Sotin}, C.,
  {Chamberlain}, M.~C., {Hansen}, G., {Hibbitts}, K., {Showalter}, M.,
  {Filacchione}, G., 2003. {Cassini-VIMS at Jupiter: solar occultation
  measurements using Io}. Icarus 166, 75--84.

\bibitem[{{Grinspoon} et~al.(1993){Grinspoon}, {Pollack}, {Sitton}, {Carlson},
  {Kamp}, {Baines}, {Encrenaz}, and {Taylor}}]{gri93}
{Grinspoon}, D.~H., {Pollack}, J.~B., {Sitton}, B.~R., {Carlson}, R.~W.,
  {Kamp}, L.~W., {Baines}, K.~H., {Encrenaz}, T., {Taylor}, F.~W., 1993.
  {Probing Venus's cloud structure with Galileo NIMS}. Plan. \& Space Sci. 41,
  515--542.

\bibitem[{{Hansen} and {Travis}(1974)}]{han74}
{Hansen}, J.~E., {Travis}, L.~D., 1974. {Light scattering in planetary
  atmospheres}. Space Sci. Rev. 16, 527--610.

\bibitem[{{Henyey} and {Greenstein}(1941)}]{hen41}
{Henyey}, L.~G., {Greenstein}, J.~L., 1941. {Diffuse radiation in the Galaxy}.
  Astrophys. J. 93, 70--83.

\bibitem[{{Hovenier} and {de Haan}(1985)}]{hov85}
{Hovenier}, J.~W., {de Haan}, J.~F., 1985. {Polarized light in planetary
  atmospheres for perpendicular directions}. Astron. Astrophys. 146, 185--191.

\bibitem[{{Hubbard} et~al.(2001){Hubbard}, {Fortney}, {Lunine}, {Burrows},
  {Sudarsky}, and {Pinto}}]{hub01}
{Hubbard}, W.~B., {Fortney}, J.~J., {Lunine}, J.~I., {Burrows}, A., {Sudarsky},
  D., {Pinto}, P., 2001. {Theory of Extrasolar Giant Planet Transits}.
  Astrophys. J. 560, 413--419.

\bibitem[{{Hubbard} et~al.(1988){Hubbard}, {Hunten}, {Dieters}, {Hill}, and
  {Watson}}]{hub88}
{Hubbard}, W.~B., {Hunten}, D.~M., {Dieters}, S.~W., {Hill}, K.~M., {Watson},
  R.~D., 1988. {Occultation evidence for an atmosphere on Pluto}. Nature 336,
  452--454.

\bibitem[{{Kim} et~al.(2012){Kim}, {Sim}, {Lee}, {Courtin}, {Moses}, and
  {Minh}}]{kim12}
{Kim}, S., {Sim}, C., {Lee}, D., {Courtin}, R., {Moses}, J., {Minh}, Y., 2012.
  {The three-micron spectral feature of the Saturnian haze: implications for
  the haze composition and formation process}. Plan. \& Space Sci. in press, in
  press.

\bibitem[{{Kim} et~al.(2011){Kim}, {Jung}, {Sim}, {Courtin}, {Bellucci},
  {Sicardy}, {Song}, and {Minh}}]{kim11}
{Kim}, S.~J., {Jung}, A., {Sim}, C.~K., {Courtin}, R., {Bellucci}, A.,
  {Sicardy}, B., {Song}, I.~O., {Minh}, Y.~C., 2011. {Retrieval and tentative
  indentification of the 3 {$\mu$}m spectral feature in Titan's haze}. Plan. \&
  Space Sci. 59, 699--704.

\bibitem[{{Kleinb{\"o}hl} et~al.(2009){Kleinb{\"o}hl}, {Schofield}, {Kass},
  {Abdou}, {Backus}, {Sen}, {Shirley}, {Lawson}, {Richardson}, {Taylor},
  {Teanby}, and {McCleese}}]{kle09}
{Kleinb{\"o}hl}, A., {Schofield}, J.~T., {Kass}, D.~M., {Abdou}, W.~A.,
  {Backus}, C.~R., {Sen}, B., {Shirley}, J.~H., {Lawson}, W.~G., {Richardson},
  M.~I., {Taylor}, F.~W., {Teanby}, N.~A., {McCleese}, D.~J., 2009. {Mars
  Climate Sounder limb profile retrieval of atmospheric temperature, pressure,
  and dust and water ice opacity}. Journal of Geophysical Research (Planets)
  114, 10006.

\bibitem[{{Krasnopol'Skii} et~al.(1989){Krasnopol'Skii}, {Moroz}, {Krysko},
  {Korablev}, and {Zhegulev}}]{kra89}
{Krasnopol'Skii}, V.~A., {Moroz}, V.~I., {Krysko}, A.~A., {Korablev}, O.~I.,
  {Zhegulev}, V.~S., 1989. {Solar occultation spectroscopic measurements of the
  Martian atmosphere at 1.9 and 3.7 microns}. Nature 341, 603.

\bibitem[{Lacis and Oinas(1991)}]{lac91}
Lacis, A.~A., Oinas, V., 1991. A description of the correlated {\em k}
  distribution method for modeling nongray gaseous absorption, thermal
  emission, and multiple-scattering in vertically inhomogeneous atmospheres. J.
  Geophys. Res. 96~(D5), 9027--9063.

\bibitem[{{Lecavelier Des Etangs} et~al.(2008){Lecavelier Des Etangs},
  {Vidal-Madjar}, {D{\'e}sert}, and {Sing}}]{lec08}
{Lecavelier Des Etangs}, A., {Vidal-Madjar}, A., {D{\'e}sert}, J.-M., {Sing},
  D., 2008. {Rayleigh scattering by H$_{2}$ in the extrasolar planet HD
  209458b}. Astron. Astrophys. 485, 865--869.

\bibitem[{{Macke} et~al.(1999){Macke}, {Mitchell}, and {Bremen}}]{mac99}
{Macke}, A., {Mitchell}, D., {Bremen}, L., 1999. {Monte Carlo radiative
  transfer calculations for inhomogeneous mixed phase clouds}. Physics and
  Chemistry of the Earth B 24, 237--241.

\bibitem[{{Miller-Ricci} and {Fortney}(2010)}]{mil10}
{Miller-Ricci}, E., {Fortney}, J.~J., 2010. {The Nature of the Atmosphere of
  the Transiting Super-Earth GJ 1214b}. Astrophys. J. Lett. 716, L74--L79.

\bibitem[{{Mishchenko} et~al.(2006){Mishchenko}, {Travis}, and {Lacis}}]{mis06}
{Mishchenko}, M.~I., {Travis}, L.~D., {Lacis}, A., 2006. {Multiple Scattering
  of Light by Particles}. Cambridge University Press.

\bibitem[{{Mishchenko} et~al.(1996){Mishchenko}, {Travis}, and {Macke}}]{mis96}
{Mishchenko}, M.~I., {Travis}, L.~D., {Macke}, A., 1996. {Scattering of light
  by polydisperse, randomly oriented, finite circular cylinders}. Appl.~Optics
  35, 4927--4940.

\bibitem[{{Niemann} et~al.(2005){Niemann}, {Atreya}, and {16
  colleagues}}]{nie05}
{Niemann}, H.~B., {Atreya}, S.~K., {16 colleagues}, 2005. {The abundances of
  constituents of Titan's atmosphere from the GCMS instrument on the Huygens
  probe}. Nature 438, 779--784.

\bibitem[{{Palmer} and {Williams}(1975)}]{pal75}
{Palmer}, K.~F., {Williams}, D., 1975. {Optical constants of sulfuric acid -
  Application to the clouds of Venus}. Appl.~Optics 14, 208--219.

\bibitem[{{Pont} et~al.(2008){Pont}, {Knutson}, {Gilliland}, {Moutou}, and
  {Charbonneau}}]{pon08}
{Pont}, F., {Knutson}, H., {Gilliland}, R.~L., {Moutou}, C., {Charbonneau}, D.,
  2008. {Detection of atmospheric haze on an extrasolar planet: the 0.55-1.05
  {$\mu$}m transmission spectrum of HD 189733b with the HubbleSpaceTelescope}.
  Monthly Not. Roy. Astr. Soc. 385, 109--118.

\bibitem[{Rothman et~al.(2009)Rothman, Gordon, and {42 colleagues}}]{rot09}
Rothman, L.~S., Gordon, I., {42 colleagues}, 2009. The {HITRAN} 2008 molecular
  spectroscopic database. J. Quant. Spectro. Rad. Trans. 96, 139--204.

\bibitem[{{Seager} and {Deming}(2010)}]{sea10}
{Seager}, S., {Deming}, D., 2010. {Exoplanet Atmospheres}. Annual Review of
  Astronomy and Astrophysics 48, 631--672.

\bibitem[{{Seager} and {Sasselov}(2000)}]{sea00}
{Seager}, S., {Sasselov}, D.~D., 2000. {Theoretical Transmission Spectra during
  Extrasolar Giant Planet Transits}. Astrophys. J. 537, 916--921.

\bibitem[{{Shettle} and {Weinman}(1970)}]{shet70}
{Shettle}, E.~P., {Weinman}, J.~A., 1970. {The Transfer of Solar Irradiance
  Through Inhomogeneous Turbid Atmospheres Evaluated by Eddington's
  Approximation.} Journal of Atmospheric Sciences 27, 1048--1055.

\bibitem[{{Sing} et~al.(2011){Sing}, {Pont}, {Aigrain}, {Charbonneau},
  {D{\'e}sert}, {Gibson}, {Gilliland}, {Hayek}, {Henry}, {Knutson}, {Lecavelier
  Des Etangs}, {Mazeh}, and {Shporer}}]{sin11}
{Sing}, D.~K., {Pont}, F., {Aigrain}, S., {Charbonneau}, D., {D{\'e}sert},
  J.-M., {Gibson}, N., {Gilliland}, R., {Hayek}, W., {Henry}, G., {Knutson},
  H., {Lecavelier Des Etangs}, A., {Mazeh}, T., {Shporer}, A., 2011. {Hubble
  Space Telescope transmission spectroscopy of the exoplanet HD 189733b:
  high-altitude atmospheric haze in the optical and near-ultraviolet with
  STIS}. Monthly Not. Roy. Astr. Soc. 416, 1443--1455.

\bibitem[{{Snellen} et~al.(2010){Snellen}, {de Kok}, {de Mooij}, and
  {Albrecht}}]{sne10}
{Snellen}, I.~A.~G., {de Kok}, R.~J., {de Mooij}, E.~J.~W., {Albrecht}, S.,
  2010. {The orbital motion, absolute mass and high-altitude winds of exoplanet
  HD209458b}. Nature 465, 1049--1051.

\bibitem[{{Stern} et~al.(2008){Stern}, {Slater}, {Scherrer}, {Stone}, {Dirks},
  {Versteeg}, {Davis}, {Gladstone}, {Parker}, {Young}, and {Siegmund}}]{ste08}
{Stern}, S.~A., {Slater}, D.~C., {Scherrer}, J., {Stone}, J., {Dirks}, G.,
  {Versteeg}, M., {Davis}, M., {Gladstone}, G.~R., {Parker}, J.~W., {Young},
  L.~A., {Siegmund}, O.~H.~W., 2008. {ALICE: The Ultraviolet Imaging
  Spectrograph Aboard the New Horizons Pluto-Kuiper Belt Mission}. Space Sci.
  Rev. 140, 155--187.

\bibitem[{{Tinetti} et~al.(2007){Tinetti}, {Vidal-Madjar}, {Liang}, {Beaulieu},
  {Yung}, {Carey}, {Barber}, {Tennyson}, {Ribas}, {Allard}, {Ballester},
  {Sing}, and {Selsis}}]{tin07}
{Tinetti}, G., {Vidal-Madjar}, A., {Liang}, M.-C., {Beaulieu}, J.-P., {Yung},
  Y., {Carey}, S., {Barber}, R.~J., {Tennyson}, J., {Ribas}, I., {Allard}, N.,
  {Ballester}, G.~E., {Sing}, D.~K., {Selsis}, F., 2007. {Water vapour in the
  atmosphere of a transiting extrasolar planet}. Nature 448, 169--171.

\bibitem[{{Tomasko} et~al.(2008){Tomasko}, {Doose}, {Engel}, {Dafoe}, {West},
  {Lemmon}, {Karkoschka}, and {See}}]{tom08}
{Tomasko}, M.~G., {Doose}, L., {Engel}, S., {Dafoe}, L.~E., {West}, R.,
  {Lemmon}, M., {Karkoschka}, E., {See}, C., 2008. {A model of Titan's aerosols
  based on measurements made inside the atmosphere}. Plan. \& Space Sci. 56,
  669--707.

\bibitem[{{Tomasko} et~al.(1999){Tomasko}, {Doose}, {Lemmon}, {Smith}, and
  {Wegryn}}]{tom99}
{Tomasko}, M.~G., {Doose}, L.~R., {Lemmon}, M., {Smith}, P.~H., {Wegryn}, E.,
  1999. {Properties of dust in the Martian atmosphere from the Imager on Mars
  Pathfinder}. J. Geophys. Res. 104, 8987--9008.

\bibitem[{{Vandaele} et~al.(2008){Vandaele}, {De Mazi{\`e}re}, {Drummond},
  {Mahieux}, {Neefs}, {Wilquet}, {Korablev}, {Fedorova}, {Belyaev},
  {Montmessin}, and {Bertaux}}]{van08}
{Vandaele}, A.~C., {De Mazi{\`e}re}, M., {Drummond}, R., {Mahieux}, A.,
  {Neefs}, E., {Wilquet}, V., {Korablev}, O., {Fedorova}, A., {Belyaev}, D.,
  {Montmessin}, F., {Bertaux}, J.-L., 2008. {Composition of the Venus
  mesosphere measured by Solar Occultation at Infrared on board Venus Express}.
  Journal of Geophysical Research (Planets) 113, 0.

\bibitem[{{Vinatier} et~al.(2006){Vinatier}, {B{\' e}zard}, and {13
  colleagues}}]{vin06}
{Vinatier}, S., {B{\' e}zard}, B., {13 colleagues}, 2006. {Vertical abundance
  profiles of hydrocarbons in Titan's atmosphere at 15$^\circ$S and 80$^\circ$N
  retrieved from Cassini/CIRS spectra.} Icarus 188, 120--138.

\end{thebibliography}

\label{lastpage}
\end{document}